# All-electrical measurements of direct spin Hall effect in GaAs with Esaki diode electrodes.

M. Ehlert[1], C. Song[1,2], M. Ciorga[1,*], M. Utz[1], D. Schuh[1], D. Bougeard[1], and D. Weiss[1]

[1]*Institute of Experimental and Applied Physics, University of Regensburg, D-93040 Regensburg, Germany*
[2]*Laboratory of Advanced Materials, Department of Material Science & Engineering, Tsinghua University, Beijing 100084, China*

Abstract

We report on measurements of direct spin Hall effect in a lightly *n*-doped GaAs channel with conductivity below 2000 $\Omega^{-1}$ m$^{-1}$. As spin detecting contacts we employed highly efficient ferromagnetic Fe/(Ga,Mn)As/GaAs Esaki diode structures. We investigate bias and temperature dependence of the measured spin Hall signal and evaluate the value of total spin Hall conductivity and its dependence on channel conductivity and temperature. From the results we determine skew scattering and side-jump contribution to the total spin Hall conductivity and compare it with the results of experiments on higher conductive *n* – GaAs channels [Phys. Rev. Lett. **105**, 156602, (2010)]. As a result we conclude that both skewness and side jump contribution cannot be treated as fully independent on the conductivity of the channel.

The spin Hall effect (SHE)[1], predicted in 1971[2], has grown to become the subject of intensive theoretical[3–6] and experimental[7-12] studies over the last decade, as one of the phenomena exploiting the electron spin degree of freedom[13]. The origin of the effect is coupling of spin and charge currents due to spin-orbit interaction in a given material.

Although electrical in nature it was initially observed optically[7,8] and fully electrical measurements of a direct SHE was observed only very recently[10].

In this paper we describe the results of the measurements of direct SHE in lightly $n$-doped GaAs channels, with spin accumulation detected by probes containing (Ga,Mn)As/GaAs Esaki diode structures[14]. The relatively high detection efficiency of the latter[15,16] allows for effective measurement of low level polarizations generated by SHE. The geometry of the measurements is similar to the one in Ref. 10. The charge current $j_x$ flows along the channel with conductivity $\sigma_{xx}$. As a result of spin-orbit interaction the scattering at impurities is spin-dependent leading to deflection of carriers with opposite spins into opposite directions, transverse to the driven charge current. This gives rise to the spin current $j_s$ in the direction perpendicular to $j_x$ with spins partially polarized in a direction perpendicular to the plane formed by $j_x$ and $j_s$. The generated spin current leads then to the spin accumulation at the edges of the channel, which is probed by, placed above the channel, ferromagnetic voltage probes with (Ga,Mn)As/GaAs Esaki diode structures. Due to spin-charge coupling occurring in the ferromagnetic materials[17] the spin accumulation in the channel leads to a voltage drop across the contact that can be experimentally measured[18]. From the measurements we extracted the value of spin Hall conductivity $\sigma_{SH}$, defined as $\sigma_{SH} = j_s / E_x$, where $E_x$ is the electric field along the channel. We investigated also its dependence on the conductivity of the channel as well as on the temperature. From these dependencies we were able to determine the contribution from skew scattering and side jump and compare it to both theoretical predictions[5] and previous experimental results[10]. In contrast to experiments by Garlid et al[10] we focused our studies on $n$-GaAs layers with relatively low carrier concentration ($n \approx 2\times10^{16}$ cm$^{-3}$) and subsequent low channel conductivities. These two experiments complement therefore each other in terms of the investigated range of channel conductivity values.

Experimental devices were fabricated from a single wafer grown by molecular-beam-epitaxy (MBE) on (001) GaAs substrates. This is one of the wafers used also for spin injection experiments described in Ref. 16. The wafer consists of a 1000 nm thick $n$-type transport channel ($n \approx 2\times10^{16}$ cm$^{-3}$, as extracted from magnetotransport measurements), a 15 nm thin $n \rightarrow n^+$ GaAs transition layer ($n^+= 5\times10^{18}$ cm$^{-3}$), 8 nm $n^+$-GaAs, and 2.2 nm low-temperature (LT)-grown Al$_{0.36}$Ga$_{0.64}$As, serving as a diffusion barrier, followed by LT-grown 15 nm thick layer of Ga$_{0.95}$Mn$_{0.05}$As. The highly doped (Ga,Mn)As/GaAs pn-junction forms an Esaki diode[14,15]. In a next step the wafers were transferred, without breaking vacuum, into an attached metal-MBE chamber, where 2 nm [14 monolayers (MLs)] of Fe were epitaxially grown at room temperature, and finally covered by 4 nm (20 MLs) of Au.

The Hall bar devices were defined by optical lithography, chemically-assisted ion beam etching and wet etching. Electron beam lithography was used to pattern the Fe/(Ga,Mn)As spin detecting contacts, oriented along the [110] direction, i.e., the easy axis of Fe. This assured that magnetizations stayed aligned along the contacts during Hanle measurements described in the next paragraph. A picture of one of the devices and the geometry used in experiments is shown in Fig.1a. A pair of 2.5 µm wide spin probes is placed at Hall crosses, with a distance $L$ from the edge of the bar. The experiments were performed on devices with $L$ = 5.25, 8.25 and 11.25 µm, measured from the center of the contact. Spin-detecting contacts are connected to big bonding pads via Ti/Au paths, which are isolated from the conducting channel by a 50 nm thick layer of Al$_2$O$_3$, deposited by atomic layer deposition (ALD). In the transport experiments the current $j_x$ is passed along the channel characterized by conductivity $\sigma_{xx}$. Ferromagnetic Esaki diodes of 100 µm × 100 µm size are also used as charge current leads, but placing them ~200 µm from the nearest pair of spin Hall probes assures that the charge current flowing underneath the probes is fully unpolarized. Due to SHE the carriers with opposite spins are accumulated at opposite edges of the channel, which can be described by spin accumulation µ$_s$. As a result of spin-charge coupling this spin accumulation gives rise to a voltage[17,18] $V = -P\mu_s$, where $P$ is spin injection efficiency of the employed contact. The feasibility of a scheme employing structures described above for studying of spin dependent phenomena has been demonstrated in non-local spin injection experiments described in Ref.16. Based on those measurements we estimated the value of $P \approx 0.5$. Additionally we performed similar measurements (not shown here) on the current devices using each of the ferromagnetic contacts as a detector of spin accumulation generated by electrical spin injection at the other contact from the pair. Although that configuration was not optimal for a spin injection experiment and could not be used for quantitative analysis, it clearly demonstrated sensitivity of contacts to the polarized spins accumulated underneath. As spin accumulation at opposite edges of the channel has opposite spin orientation, the voltage between both contacts $V_{SH}$ can be written as $|V_{SH}| = 2P\mu_s$, assuming the same spin injection efficiency for both contacts. During measurements we were also monitoring the voltage $V_{cd}$ resulting from the ordinary Hall effect, as this produced background to the spin Hall signal.

Ferromagnetic electrode can measure only spin components parallel to its own magnetization axis, which for our structure lies in the plane of the sample. As the SHE-induced spin polarization is aligned along the $z$-direction we apply external magnetic field $\mathbf{B}_y$ to induce precession of the spins in $xz$ plane due to Hanle effect. As a result of the precession spins acquire an in-plane component that can be detected by spin-dependent contacts. Typical results of measurements are shown in Fig.1b, where we plot the voltage $V_{ab}$ measured between a pair of contacts in the applied external magnetic field $\mathbf{B}_y$. The following procedure was applied to obtain the shown curves. First magnetic field $\mathbf{B}_x$ was swept to the saturation value of 0.5 T and back to zero to align the magnetization along the contacts in $+x$-direction. Then the sample was rotated in-plane by 90 deg and the field $\mathbf{B}_y$ was swept from zero to 0.5 T to induce precession of the out-of plane spin component. The procedure was then repeated with $\mathbf{B}_y$ swept from zero to -0.5 T. Although the raw curves, shown in Fig 1b, contain contributions from different backgrounds, they clearly show the features expected from a spin-related signal, namely antisymmetric behavior near $\mathbf{B}_y$=0 T and opposite sign of the signal for magnetizations aligned along +x and –x direction. To be able to fit the data using standard Hanle equations[18] we had to remove the background contribution. First we removed the background due to ordinary Hall effect by subtracting the Hall voltage $V_{cd}$ from measured $V_{ab}$. The remaining background was removed by taking advantage of the expected symmetry of the spin Hall signal. This was done by subtracting the curves taken for two different parallel configurations (magnetized at + x and –x) and subsequently removing the even components from the data as the expect signal should be odd in $\mathbf{B}_y$. The traces obtained after removing the background are depicted in Fig. 1c. We plot the data for two opposite current directions, clearly showing that the sign of the signal is changed by reversal of current direction, which is fully consistent with the theory of SHE

Figures 2(a)–(c) show the curves for $j_x$ = 1.7×10$^3$ A/cm$^2$ and T = 4.2 K obtained for different distances of contacts from the channel edge, after removing all contributing backgrounds. For fitting we used standard Hanle effect equations[15,18] for the case of perpendicular relative orientation of spins (in our case originating from SHE) and spin-detector. We took the final size of the contacts into account by integrating the signal over their width. From the fits we obtained a spin relaxation time of $\tau_s$ = 3.5 ns and $V_0$ = 83μV, where $V_0$ is the voltage corresponding to the spin accumulation at the edges of the contact. This voltage is directly related to the spin density polarization $P_n = (n^\uparrow - n^\downarrow)/(n^\uparrow + n^\downarrow)$ through the expression[17]

$$P_n = \frac{eV_{SH}}{2P}\frac{g_n(E_F)}{n} = \frac{eV_{SH}}{2P}\frac{3m^*}{\hbar^2(3\pi^2 n)^{2/3}} \quad (1)$$

where $g_n(E_F)$ is the density of states at the Fermi energy and $m^*$ is the effective mass of GaAs. From the measurements shown in Fig.2 we obtain $P_n^0 = 3\%$ at the edges of the sample, which is roughly double the value of the spin polarization obtained for higher doped n-GaAs.[10]

Now we can move on to extracting the parameters related directly to SHE, namely the magnitude of the spin Hall conductivity $\sigma_{SH}$ and spin Hall angle $\alpha = \sigma_{SH}/\sigma_{xx}$. As mentioned before a transverse spin current $j_s$ is creating at the edges of the channel a spin accumulation $\mu_s = j_s \rho_n \lambda_{sf}$ which via spin charge coupling leads to voltage $|V_0| = 2P\mu_s$. Taking into account that $j_s = \sigma_{SH} E_x$ and $j = \sigma_{xx} E_x$ we derive

$$\sigma_{SH} = V_0 \sigma_{xx}^2 / 2Pj\lambda_{sf} \quad (2)$$

From our measurements we obtained $V_0 = 83$ µV for $j = 1.7 \times 10^3$ A/cm$^2$, $\lambda_{sf} = 8.5$ µm (Fig.2d) and $\sigma_{xx} = 1370$ $\Omega^{-1}$m$^{-1}$. This finally gives $\sigma_{SH} = 1.08$ $\Omega^{-1}$m$^{-1}$ and $\alpha = 7.8 \times 10^{-4}$, i.e., values which are very consistent with other reports[5,10].

To analyze the obtained SHE signal in more details and to determine the contribution of side jump and skew scattering to the measured spin Hall conductivity we performed bias dependence measurements for current densities in the range of $j = 3.3 \times 10^2 - 3.3 \times 10^3$ A/cm$^2$ at $T = 4.2$K. In Fig 3(a)–(c) we show the spin Hall signal (symbols) for three different values of current density together with Hanle fits (solid lines) from which we can extract the spin Hall conductivity, in the way described above. Changing the current density tunes the value of the conductivity $\sigma_{xx}$[10] in $n$-GaAs due to dependence of mobility on the electric field [16]. This allows us to extract dependence of $\sigma_{SH}$ on $\sigma_{xx}$, plotted in Fig. 3(d) as filled squares. According to Engel *et al*[5] $\sigma_{SH}$ can be approximated by

$$\sigma_{SH} \approx \frac{\gamma}{2}\sigma_{xx} + \sigma_{SJ} \quad (3).$$

The first component in the above equation is due to skew scattering with $\gamma$ being the so-called skewness parameter. The second component describes side jump contribution. According to theory it is independent on the conductivity of the channel and depends on density $n$ and spin orbit interaction parameter $\lambda_{so}$ as

$$\sigma_{SJ} = -2ne^2\lambda_{so}/\hbar \quad (4).$$

From a linear fit of our data with Eq. (3) we obtained the skewness parameter $\gamma = 4.3 \times 10^{-4}$ and the conductivity-independent side jump contribution $\sigma_{SJ} = 0.58 \times 10^{-1} \, \Omega^{-1} m^{-1}$. The first value compares quite well to the value $\gamma \approx 1/900$ calculated in Ref. 5 and is approximately one order of magnitude smaller than the value obtained in Ref. 10. The value of the side jump contribution differs by $\sim 1.1 \times 10^{-1} \Omega^{-1} m^{-1}$ from the one predicted by Eq.4 (the fact that the experimental value has a positive sign is an artifact of linear extrapolation). This difference between theory and experiments is approximately one order of magnitude smaller than the one reported by Garlid *et al.*[10] In order to make a more detailed comparison of our experimental results and the ones from Ref. 10 we plot in Fig 3(d) also the dependence of the spin Hall conductivity on channel conductivity reported in that paper for T = 30 K (open symbols). One can see that the much higher values for skewness and side jump contribution, given in the latter, were extracted from spin Hall signals measured for conductivities larger than $\sim 3000 \, \Omega^{-1} m^{-1}$ (see the linear fit). In contrast, as a consequence of a lower doping, we performed measurements on channels with conductivities up to $\sim 1600 \, \Omega^{-1} m^{-1}$. Calculations of Engel *et al* were performed for channels with parameters similar to our samples, what can explain why our results are closer to their theoretical predictions. One clearly sees however that data from Ref. 10 contain also values of $\sigma_{SH}$ extracted for the conductivity range of $\sim 2500-3000 \, \Omega^{-1} m^{-1}$. The corresponding data points deviate substantially from the extrapolated line from which $\gamma$ and $\sigma_{SJ}$ were extracted and are closer to our results. Our experiment provides then additional data points for channel conductivities below 2000 $\Omega^{-1} m^{-1}$, which seem to fit well with data points for $\sigma_{xx} \approx 2500-3000 \, \Omega^{-1} m^{-1}$ from Ref. 10. Both experiments complement each other well and together show that spin Hall conductivity can be well described by Eq. 3, however one cannot treat both skewness parameter $\gamma$ and side jump contribution $\sigma_{SJ}$ as fully independent on $\sigma_{xx}$. There seem to exist two regimes in $\sigma_{xx}$ in which two different sets of parameters $\gamma$ and $\sigma_{SJ}$ determine spin Hall conductivity $\sigma_{SH}$.

We performed also temperature studies of the spin Hall signal in the range of T = 4.2–80 K for $j = 1.7 \times 10^3$ A/cm$^2$. The results are summarized in Fig.4, with figures (a)–(c) showing experimental data (symbols) and Hanle fits (solid lines) for three different temperature values. In Fig. 4(d) we plot then the dependence of extracted spin Hall conductivity on temperature. The measured signal decreased with increasing temperature mainly as a result of decreasing $\tau_s$. Above T = 70 K the signal was no longer observable, which is consistent with spin injection experiments on the same wafer.[16] Because channel conductivity increases with temperature we expected also an increase of $\sigma_{SH}$. This was indeed observed as shown in Fig. 4(d). In the same figure we plot also the predicted dependence (red triangles) of $\sigma_{SH}$ on temperature using Eq. 3 and the measured values of $\sigma_{xx}(T)$ and n(T). We clearly see that extracted $\sigma_{SH}$ increases faster with T than predicted, especially in the range of 4–30K.

In summary, we conclusively demonstrated all-electrical measurements of spin Hall effect using Esaki diodes as ferromagnetic spin detectors. The high spin detection efficiency of the latter results in a relatively high amplitude of the measured signal, comparing, e.g., to experiments with Fe/GaAs Schottky diodes as spin sensitive contacts.[10] This allowed us to efficiently study spin Hall effect in channels with lower conductivities than previously. The values of spin Hall conductivities extracted from our measurements are consistent with those calculated by Engel *et al*[5] and smaller than those presented in Ref. 10. Combined results of these two experiments show that both skewness and side jump contribution into spin Hall conductivity can be treated as channel conductivity-independent only in a certain regime of the latter and may have different values in different ranges of conductivity.

This work has been supported by the Deutsche Forschungsgemeinschaft (DFG) via SFB689 and SPP1285 projects. C.S. is grateful for the support of Alexander von Humboldt Foundation.

*corresponding author: mariusz.ciorga@physik.uni-regensburg.de

**References**

1. J. E. Hirsch, Phys. Rev. Lett. **83**, 1834 (1999).

2. M. I. D'yakonov and V. I. Perel, Phys. Lett. **35A**, 459 (1971).

3. S. Murakami, N. Nagaosa, and S.-C. Zhang, Science **301**, 1348 (2003).

4. J. Sinova *et al.*, Phys. Rev. Lett. **92**, 126603 (2004).

5. H. A. Engel, B. I. Halperin, and E. I. Rashba, Phys. Rev. Lett. **95**, 166605 (2005).

6. W. K. Tse and S. Das Sarma, Phys. Rev. Lett. **96,** 056601 (2006).

7. Y. K. Kato, R. C. Myers, A. C. Gossard, and D. D. Awschalom, Science **306**, 1910 (2004).

8. J. Wunderlich, B. Kaestner, J. Sinova, and T. Jungwirth, Phys. Rev. Lett. **94**, 047204 (2005).

9. S. O. Valenzuela and M. Tinkham, Nature **442**, 176 (2006).

10. E. S. Garlid et al., Phys. Rev. Lett. **105**, 156602 (2010).

11. K. Ando and E. Saitoh, Nature Comm. **3**, 629 (2012).

12. T. Jungwirth, J. Wunderlich, and K. Olejník, Nature Mater. **11**, 382 (2012).

13. I. Žutić, J. Fabian and S. Das Sarma, Rev. Mod. Phys. **76**, 323 (2004).

14. M. Kohda *et al*, Jpn. J. Appl. Phys., Part 2 **40**, L1274 (2001) ; E. Johnston-Halperin *et al.*, Phys. Rev. B **65**, 041306R(2002) ; P. Van Dorpe *et al.*, Appl. Phys. Lett. **84**, 3495 (2004).

15. M. Ciorga *et al.*, Phys. Rev. B **79**, 165321 (2009).

16. C. Song *et al.*, Phys. Rev. Lett., **107**, 056601 (2011).

17. M. Johnson and R. H. Silsbee, Phys. Rev. Lett. **55**, 1790 (1985).

18. J. Fabian *et al.* Acta Phys. Slov. **57**, 565 (2007).

19. D. J. Oliver. Phys. Rev. **127**, 1045 (1962).

**Figure captions**

**Figure 1**. (color online) (a) micrograph of a spin Hall device with experimental layout. A pair of ferromagnetic contacts (*a* and *b*) is placed at the Hall cross at a distance *d* from the channel edges. Voltage $V_{ab}$ and Hall voltage $V_{cd}$ is measured as a function of in-plane field $\mathbf{B}_y$, in the presence of charge current $j_x$. (b) $V_{ab}$ as a function of $\mathbf{B}_y$ for $j_x = 1.7 \times 10^3$ A/cm$^2$ for initial orientation of magnetizations along +*x* (solid black symbols) and –*x* (open red symbols) directions. (c) Spin Hall voltage $V_{SH}$ vs. $\mathbf{B}_y$ for the same current amplitude for both positive (solid black symbols) and negative (open red symbols) current directions after removing the background as described in the text.

**Figure 2**. (color online) (a) – (c) spin Hall voltage $V_{SH}$ as a function of $\mathbf{B}_y$ at T = 4.2 K and for $j_x = 1.7 \times 10^3$ A/cm$^2$ for three different values of a distance between contact and the channel edge. Solid lines represent fits with the same set of parameters. (d) Magnitude of the spin Hall signal $\Delta V_{SH}$ vs. distance between contact and the channel edge with the extracted value of the spin diffusion length $\lambda_{sf}$.

**Figure 3**. (color online) (a) – (c) spin Hall voltage $V_{SH}$ as a function of $\mathbf{B}_y$ at T = 4.2 K for three different channel current densities $j_x$ corresponding to different conductivities $\sigma_{xx}$. We plot both experimental data (symbols) and corresponding fits (red line); (d) dependence of the spin Hall conductivity on $\sigma_{xx}$ extracted from our measurements at T=4.2K (full symbols) and taken from Ref. 10 (open symbols, T = 30 K). From the linear interpolation (solid line) with the equation (3) we obtain the value of skewness parameter $\gamma = 4.3 \times 10^{-4}$ and side jump contribution $\sigma_{SJ} \approx 0.58 \times 10^{-1} \, \Omega^{-1} \mathrm{m}^{-1}$. From the linear fit shown for Ref. 10 data one gets $\gamma = 4.0 \times 10^{-3}$ and $\sigma_{SJ} \approx -12 \times 10^{-1} \, \Omega^{-1} \mathrm{m}^{-1}$

**Figure 4** (color online) (a)–(c) spin Hall voltage $V_{SH}$ as a function of $\mathbf{B}_y$ at three different temperature values T for current density $j_x = 1.7 \times 10^3$ A/cm$^2$; (d) temperature dependence of $\sigma_{SH}$ for $j_x = 1.7 \times 10^3$ A/cm$^2$. Displayed are the experimental results (black squares) and predicted values (red triangles), derived from Eq. 3 using measured n(T) and $\sigma_{xx}$(T). Lines are just guides for the eye.

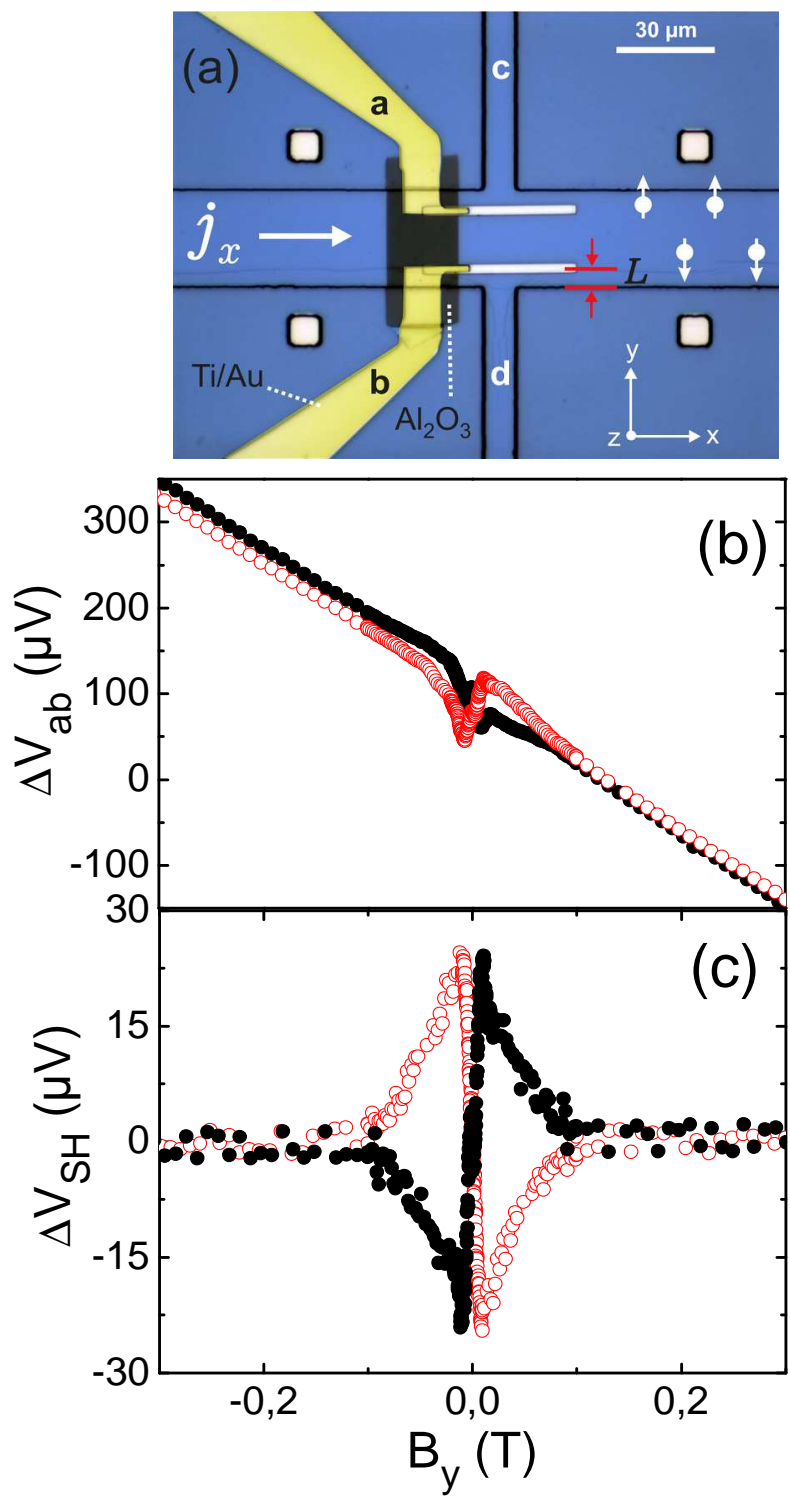

Figure 1

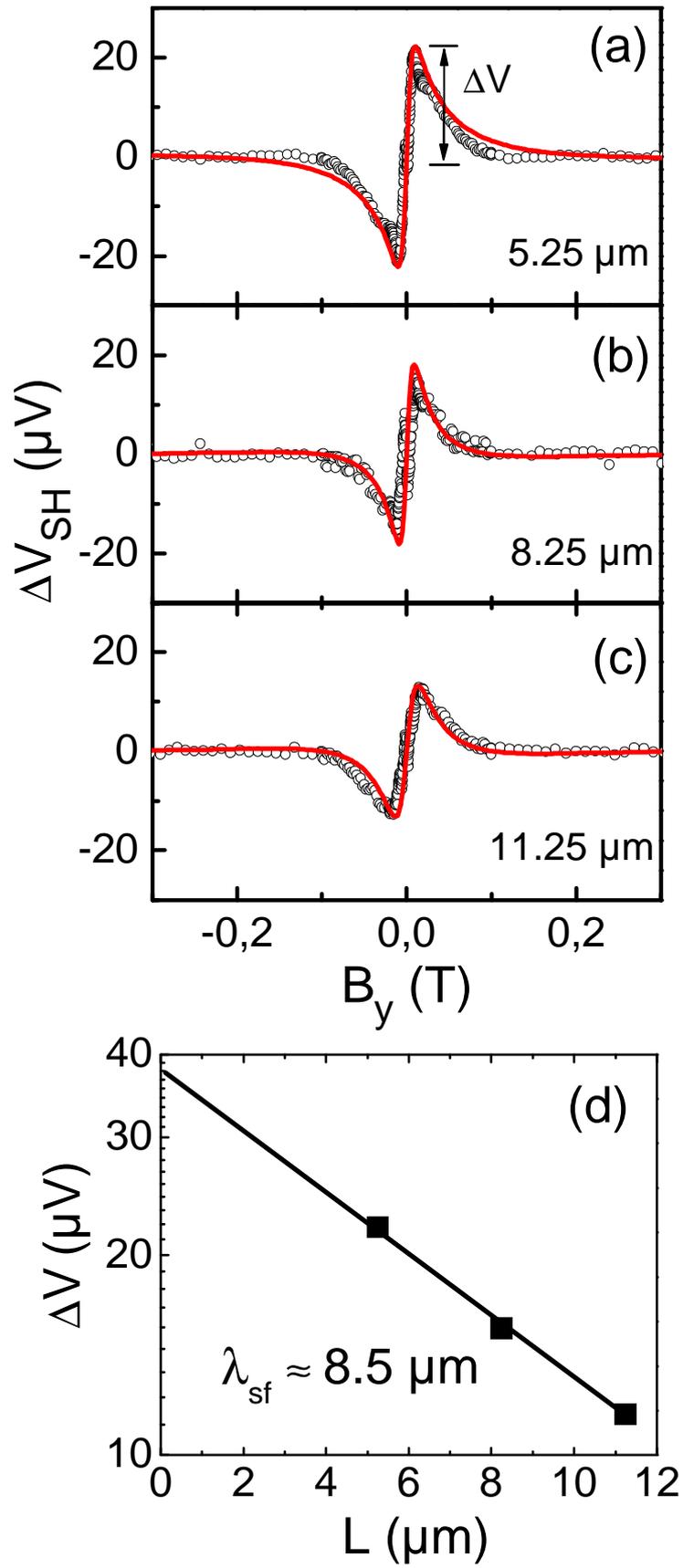

Figure 2

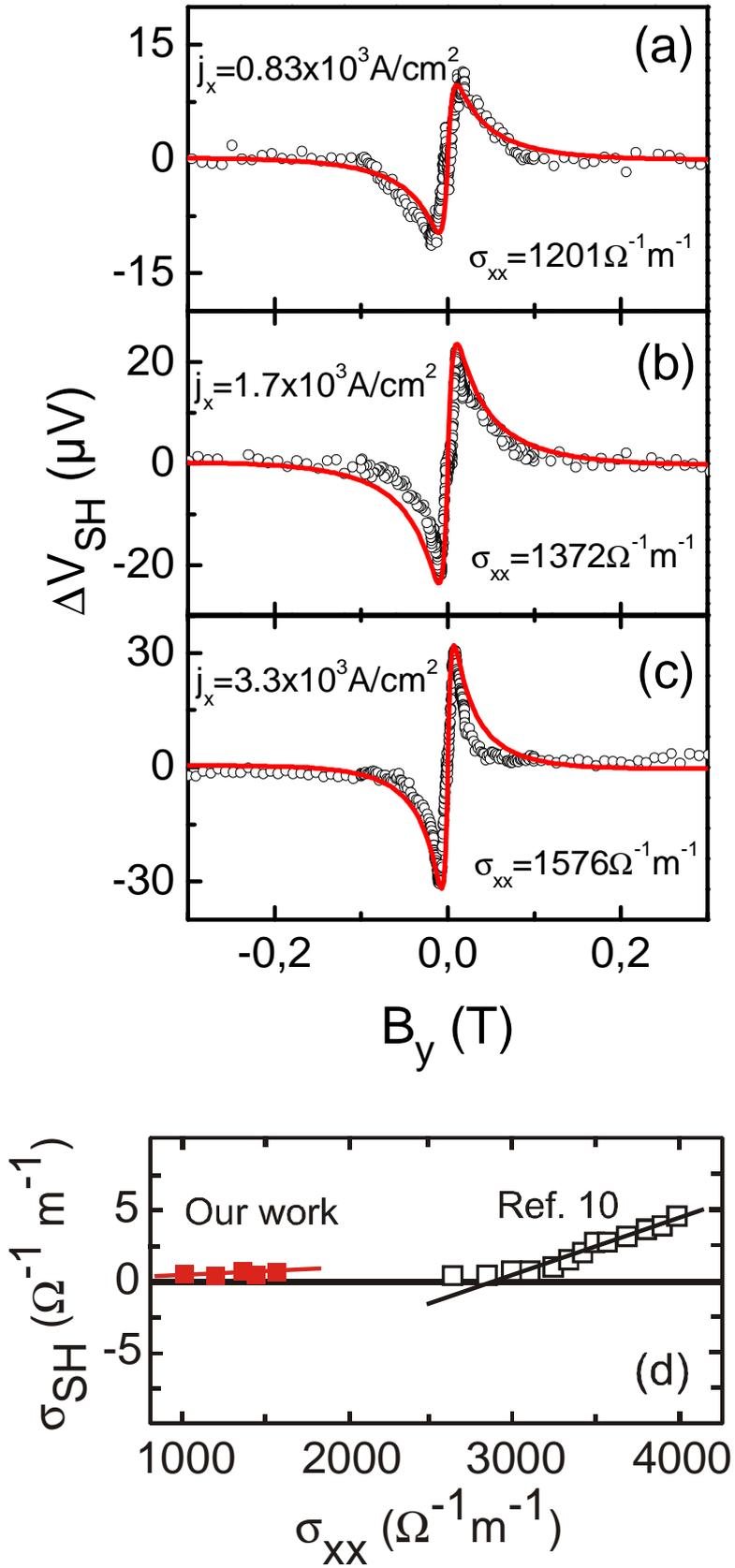

Figure 3

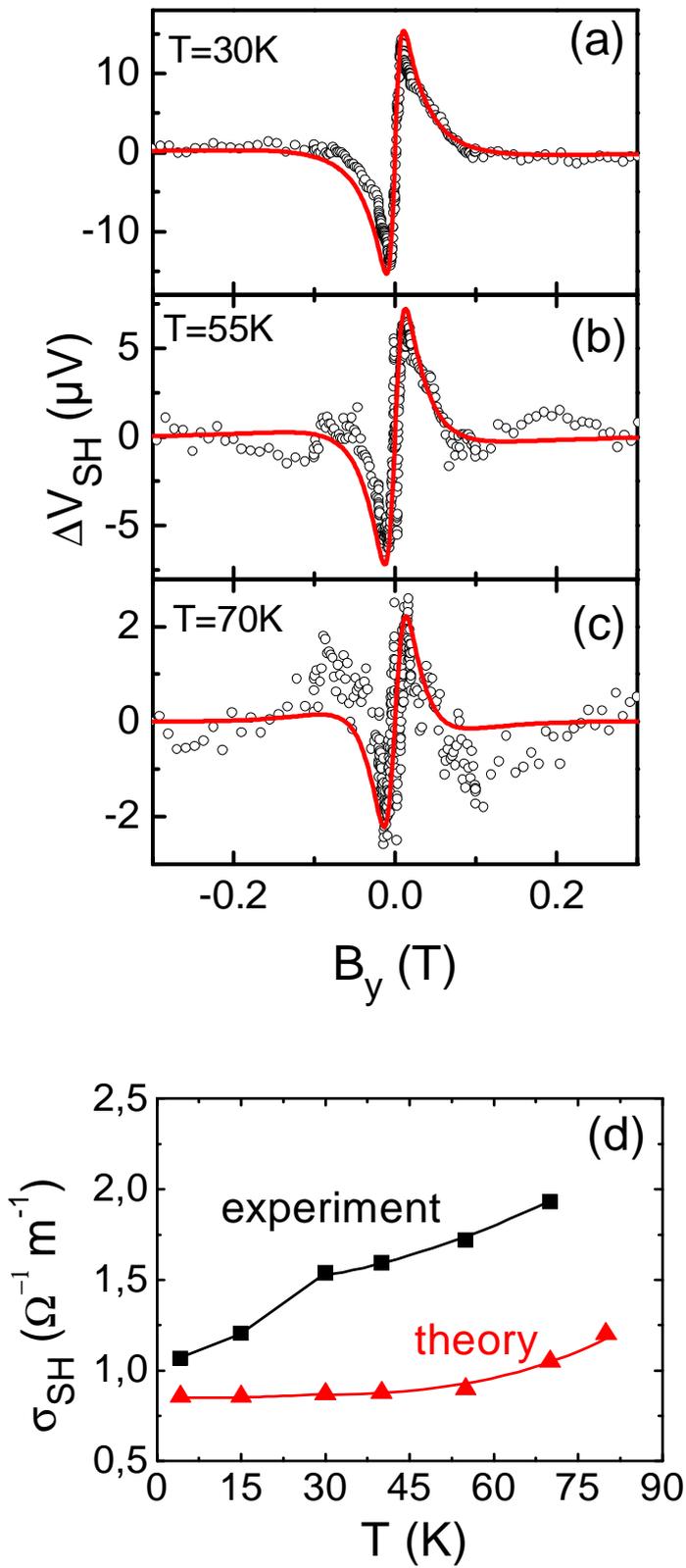

Figure 4